# Influence of local surface defects on the minority-carrier lifetime of passivating-contact solar cells

Jean Cattin[1], Jan Haschke, Christophe Ballif and Mathieu Boccard

*Photovoltaics and Thin-Film Electronics Laboratory (PV-lab), Institute of Microengineering (IMT), École Polytechnique Fédérale de Lausanne (EPFL), Rue de la Maladière 71b, CH-2002 Neuchâtel, Switzerland.*

Unlocking the full potential of passivating contacts, increasingly popular in the silicon solar cell industry, requires determining the minority carrier lifetime. Minor passivation drops limit the functioning of solar cells, however, they are not detected in devices with open-circuit voltages below 700 mV. In this work, simulations and experiments were used to show the effect of localized surface defects on the overall device performance. Although the defects did not significantly affect lifetime measurements prior to electrode deposition or open-circuit voltage measurements at standard-test conditions, it had a significant impact on the point of operation and, in turn, device efficiency (up to several percent efficiency drop). Furthermore, this study demonstrates that localized defects can have a detrimental effect on well-passivated areas located several centimeters away through electrical connection by the electrode. This leads to a low-injection lifetime drop after electrode deposition. Thus, commonly measured lifetime curves before metallization (and therefore internal voltage) are usually not representative of their respective values after metallization. The low-injection lifetime drop often observed after electrode deposition can derive from such local surface defects, and not from a homogeneous passivation drop.

Good surface passivation is essential in monocrystalline silicon solar cell technologies, as it enables taking full advantage of high-purity wafers.[1,2] Lifetime spatial non-uniformity is common in multicrystalline silicon (due to bulk variations),[3,4] and for monocrystalline devices with local contacts between the wafer and metallization.[5] Parasitic effects, such as shunts, localized recombination centers or edge recombinations are known to especially affect the low-injection (typically below a minority carrier density of $10^{15}$ cm$^{-3}$) lifetime of silicon-based devices,[3–7] as well as for undiffused monocrystalline silicon wafers passivated with charged dielectrics.[8] The interconnection of such defects with the remaining (well-passivated) device area can furthermore strengthen their detrimental influence.[5,9] The lifetime of solar-cell devices with passivating contacts (consisting of a continuous stack separating the metal electrode from the wafer) are usually characterized in terms of homogeneous lifetime, using a unique recombination parameter ($J_0$) value.[10] Surface passivation is typically followed by monitoring the effective lifetime via photoconductance decay measurements,[11] assuming that passivation occurs homogeneously.

In silicon solar cells, passivation is typically achieved by applying intrinsic amorphous silicon or silicon oxide thin films [10,12,13], whereas carrier selectivity usually relies on the work-function of a layer of doped silicon[14] or other material.[10] In such thin films, spatial passivation non-uniformities are likely to occur (e.g. a speck of dust, scratch, or inhomogeneous wet-chemistry). This can be relevant, as recombination in high-efficiency devices is limited by the surface (besides inherent Auger recombination).

This study shows that small defects in passivating contact stacks can cause performance drops of up to a few percent, therefore it may prevent future data misinterpretion and incorrect optimization procedures. Additionally, the influence of defect interconnecting through the electrode is discussed. Although this work focuses on silicon heterojunction solar cells (and uses the corresponding terminology), these observations also apply to other passivating-contact technologies, including tunnel-oxide-based and dopant-free approaches.

A schematic of the used device can be seen on Fig. 1a, and Fig. 1b shows its modelization as an array of parallel elements connected through a sheet resistance $R_{ITO}$, similarly to models of multi-crystalline devices or homojunctions cells with localized contacts.[15–19] Each element includes a current generator, three diodes to account for bulk, surface and Auger recombination, and a series resistance $R_S$ accounting for the contact resistance between the conductive sheet and the absorber. A one-dimensional array (parallel to the solar cell plane) is assumed for simplicity. This one-dimensional nature describes an infinite linear defect and would not accurately describe a point defect with a radial current spreading symmetry, which would require a two-dimensional finite-element treatment.[18,19] However, similar trends and conclusions are expected. Furthermore, as the solar cell thickness is aggregated in the diode model, diffusion within the c-Si bulk is not accounted for, making this approach inaccurate for sub-millimeter-scale analysis.

Most elements in the array have "good" properties, which

[1] Jean.cattin@epfl.ch

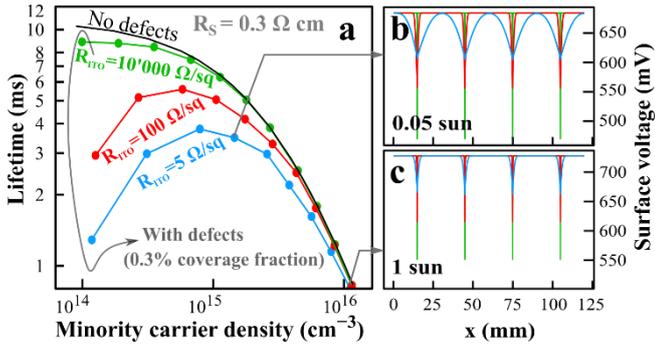

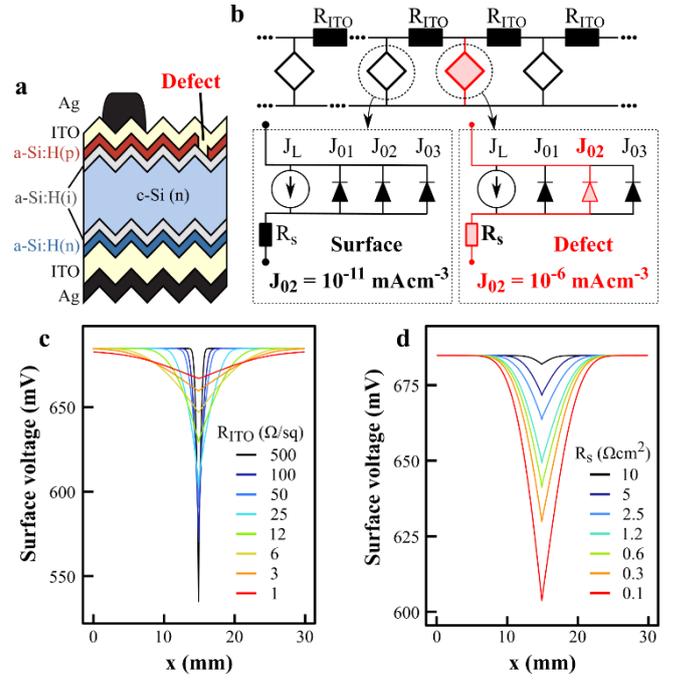

FIG. 2 a) Lifetime curves simulated from the 1D array with various sheet resistances for the interconnection, representative of the state prior to any electrode deposition (10'000 Ω/sq, corresponding to the Rsheet of the inversion layer), after ITO (100 Ω/sq), and after metal grid (5 Ω/sq, reasonable estimate since grid is not a homogeneous sheet). b,c) Surface voltage at two illumination levels for the three different $R_{ITO}$ values.

were obtained by fitting the current-voltage (JV) characteristic of an experimental solar cell with an open-circuit voltage, $V_{oc}$, of ~730 mV. One or a few elements in the array have "defect" properties, which were obtained by fitting the JV curve of a poor device showing ~540 mV of $V_{oc}$. This is considered as representative of a zone with very poor passivation, possibly resulting from the presence of dust on the surface during deposition, scratches or inhomogeneous film-deposition during processing, broken pyramid...

Fig. 1c and d show the impact of sheet resistance ($R_{ITO}$) and contact resistance ($R_S$) on the local surface voltage around a localized defect. A photocurrent ($J_L$) corresponding to an illumination of 5 mW/cm² (0.05 sun) was chosen, to be representative of the injection level at maximum power point (MPP) in standard test conditions.[20] Decreasing $R_{ITO}$ distributes the detrimental influence of the defect over a larger area and reduces the local voltage drop. Decreasing $R_S$ enlarges the range of influence of the defect, as well as the local surface voltage drop, as more charges can flow through the defect.[21] Although the device is at open-circuit, in both cases, a current flows from the good area through the conductive sheet towards the defect. The diodes' properties of the defect dictate the local potential. The local absence of p-aSi:H layer could be such defect, which is well represented by a diode (since the i-aSi:H / ITO stack acts as a Schottky contact) with a very high dark saturation current $J_{02}$.

Fig. 2 presents the impact of a homogeneous distribution of localized defects with a small coverage fraction of 0.3%. Fig. 2a shows a simulation of the lifetime curve that would be measured for such a device. The data was extracted from surface voltage simulations considering variable-illuminations (hence variable photocurrents), at open circuit, as is done in "Suns-Voc" measurements.[22] The average voltage over the entire array was determined (as commonly done) to obtain a unique value, despite spatial variations.[23,24] Fig. 2b and c show the spatial voltage variation for 0.05-sun (~MPP) and 1-sun illumination, respectively. Three values are compared for $R_{ITO}$, which are representative of a device without any electrodes (then the inversion layer at the p-n interface still leads to a conducting sheet of ~10'000 Ω/sq,[25,26]

FIG. 1 a) Schematic of the devices studied here consisting of a crystalline silicon absorber, intrinsic, p-type and n-type amorphous silicon for the contacts and indium tin oxide and silver for the electrode. b) Equivalent circuit for the simulations of localized defects, only $J_{02}$ is changed between good areas and defect. c,d) Simulated effect on the surface voltage at 0.05 sun of a single local defect as a function of the c) surface sheet resistance ($R_{ITO}$) d) series resistance ($R_s$). $R_S = 0.3$ Ω cm² in c) and $R_{ITO} = 10$ Ω in d) with only a transparent conducting layer (for which 100 Ω/sq is a usual value[27]), and after metallization (5 Ohm/sq.[2]

Fig. 2a shows that the introduction of defects has a negligible effect on the lifetime curve, when assuming a high $R_{ITO}$ value (i.e. prior to any electrode deposition), implied Voc (iVoc) and implied fill-factor (iFF) values equivalent to the no-defect case (728 mV and 87.3% respectively). However, there is a severe lifetime drop, especially at low injection, when decreasing the sheet resistance of the electrode. Hence, iFF drops to 86.8% (resp. 84.6%) for $R_{ITO}$ of 100 Ω/sq (resp. 5 Ω/sq). Nonetheless, iVoc is almost unaffected (727 mV and 725 mV, respectively). This phenomenon will more strongly affect device performance at low-injection (FF under low irradiation) and, in turn, the energy yield.[28] In these simulations, only $R_{ITO}$ was changed, with the lifetime drop deriving solely from the lateral spreading of the influence of the defects, as discussed in Fig. 2 and shown in Fig. 3b and c.

Experimental observations of such curves during operation could be misinterpreted, e.g. as passivation deterioration introduced by ITO deposition (sputter damage[29,30]) or metallization, when in this case these steps actually only reveal some pre-existing passivation issues. Moreover, as increasing $R_{ITO}$ or $R_S$ mitigates the spreading of local passivation non-uniformity (Fig. 1c and d), applying an electrode with a high sheet resistance or yielding a high contact resistance would limit the lifetime drop shown in Fig. 2a. This could be misinterpreted as improved electrode requirement for fingers in a 160-mm-wide solar cell device using 18 smartwires for interconnection, and a value of 0.3 Ωcm² is in a reasonable range.

processing, only requiring fine tuning to decrease series resistance. However, such fine tuning would annihilate the advantage of this electrode, as the poor electrical connection is the exact cause of the reduced lifetime drop. Finally, fitting

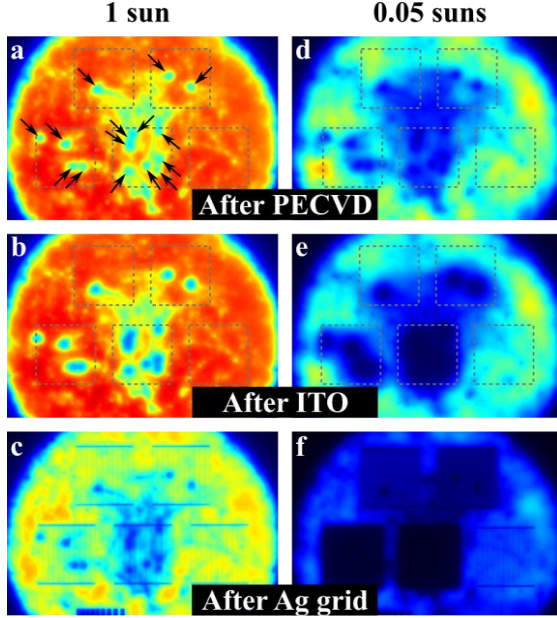

**1 sun**  **0.05 suns**

After PECVD / After ITO / After Ag grid

FIG. 4 Photoluminescence images at 1 sun (a-c), and 0.05 sun (d-e) of 2x2 cm² solar cells' wafer prior to any electrode deposition (a,d), after ITO (b,e), and after metal grid (c,f). Specially engineered localized defects are indicated by arrows in a, and dash grey squares illustrate the position of the solar cells.

the lifetime curves of Fig 2a before and after ITO deposition (implicitly assuming a homogeneous passivation) would suggest variations in surface defect density and/or fixed charge.[31] The phenomenon described here can provide alternative explanations to the experimental observations of a lifetime drop at low injection upon deposition of the transparent electrode.[31–33] This also highlights the importance of ensuring good sample homogeneity and accounting for changes in minority-carrier sheet conductance when fitting lifetime curves.

To validate these simulation findings, 2x2 cm² silicon heterojunction solar cells were fabricated using float-zone, 180-µm-thick, n-type (2 Ωcm) silicon wafers, textured by alkaline etching, as depicted in Fig. 1a. Prior to plasma-enhanced chemical vapor deposition of the a-Si layers, an HF solution was used to remove the native oxide from the surfaces. The front indium tin oxide (ITO) and rear ITO/Ag layers were deposited by means of reactive sputtering, and the front Ag grid using screen-printing. 0, 1, 2, 4, or 8 defective spots were achieved by introducing 1-mm² masks during the a-Si(p) layer deposition. Fig. 3 shows photoluminescence images illustrating the impact of illumination and electrode sheet resistance on the range of influence of the local defects. At 1-sun illumination (Voc) (Fig. 3a-c), the defects can be seen individually on PL images. The addition of a 90 Ω/sq ITO layer (Fig. 3b), followed by the screen-printed metallization (Fig. 3c), enlarges slightly the area of influence of each local defect, although this effect is negligible, as can be expected from Fig. 1. At 0.05-sun illumination (~MPP), individual defects can also be observed prior to the deposition of electrodes (Fig. 3d). However, when introducing a conducting electrode (Fig. 3e-f), the area of influence of each defect is enlarged, finally spreading to the entire cell area after metallization (Fig. 3f). Notably, the no-defect cell (low-right) does not shows any strong sputter-damage, as the area within and outside the cell are similar in color. Furthermore, although the 4-defect-cell (low-left) is only locally impacted at 1-sun illumination, the whole cell area is impacted at MPP, as seen in (Fig. 3c,f). Table 1 confirms these findings with open-circuit voltage and fill factor (FF) values for each device.

*Table 1: Voc and FF measured on the solar cells shown in Fig. 3.*

| #defects | 0 | 1 | 2 | 4 | 8 |
|---|---|---|---|---|---|
| Voc (mV) | 723 | 719 | 716 | 714 | 697 |
| FF (%) | 76.7 | 75.2 | 74.5 | 73.8 | 72.8 |

Next, the influence of local surface defects was analyzed for defects located outside the probed area (and much further than the diffusion length). Tests showed that they also can affect lifetime measurements, due to defect interconnection through the electrode. Fig. 4a depicts the sample designed to probe such a scenario through the use of an annular mask during the deposition of the a-Si:H(p) layer. The lifetime was measured in the center of this wafer, as well as in the center of a reference un-masked wafer, prior to and after deposition of a 480 Ω/sq ITO layer (Fig. 4b). Almost no lifetime modification was observed upon ITO deposition for the unmasked sample, however, a severe drop at low injection was seen for the masked sample. This is similar to the observations of Tomasi et al,[32] although in our experiment this phenomenon only occurred when the annular mask was used. We thus attribute this effect to the interconnection of the probed area (still well passivated) with the remote low-passivation annular area.

This phenomenon can be reproduced with an equivalent

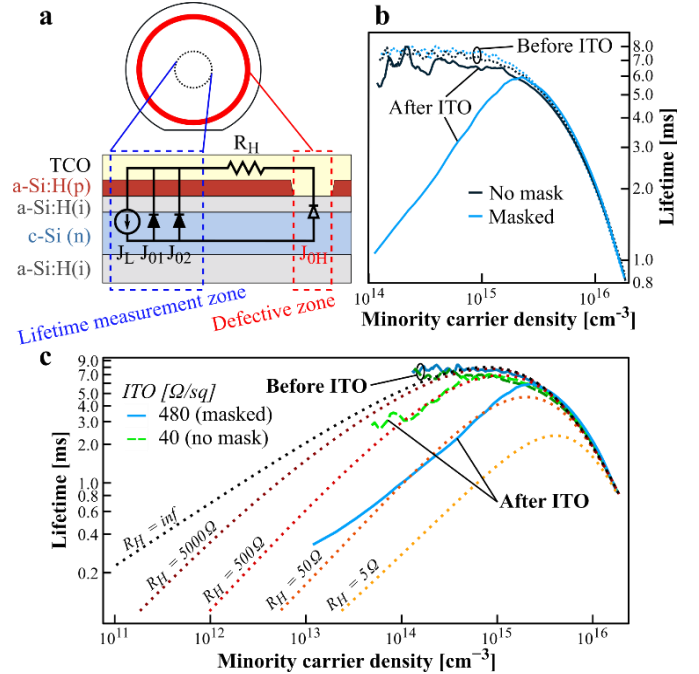

FIG. 3a) top- and side-view sketch of a wafer prepared with a tailored, annular defective area. An equivalent circuit is overlaid on the side view. b) Lifetime measured on such sample and an unmasked reference sample before and after deposition of a 480 Ω/sq ITO film. c) Same data as b overlaid with lifetime curves simulated with the 3-diode model shown in a), for various values of R_H. Green lines additionally shows experimental data before / after ITO for an unmasked sample for an ITO Rsheet of 40 Ω/sq.

circuit (overlaid on Fig. 4a) including a resistance-limited high-recombination diode. Such kind of model was previously used to explain discrepancies in current-voltage behaviors of solar cells.[9] Varying the resistance value enables reproducing the low-injection lifetime drop, as shown in Fig. 4c. In this graph, lifetime curves shown in Fig. 4b are overlaid, along with the measured lifetime of a third wafer, without any mask, before and after deposition of a 40-Ω/sq ITO. In that case, a noticeable lifetime drop is still observed, despite the absence of a mask.The ITO workfunction was expected to be fully screened, as the a-Si(p) layer was four times the standard thickness. Although this phenomenon could stem from the presence of localized defects within the probed area, it could also originate from the connection of undesired poorly-passivated areas on the wafer's periphery, be it the sample edge or tweezer marks from handling.[8,34] This effect is similar to the one previously observed for homojunction devices with a highly conductive minority-carrier collector.[9,35] Our study shows that it also applies to passivating contacts and should be considered when analyzing lifetime curves.

In conclusion, in this work we showed the impact of local surface defects on the lifetime of silicon solar cells at various processing stages. It was observed that local surface defects have a insignificant effect on lifetime prior to electrode deposition or on the Voc of a finished device. Simultaneously, the defects can affect the performance of high-efficiency devices through a reduction in internal voltage at maximum power point. The interconnection of well-passivated and defective areas through the electrode leads to a lifetime drop at low injection, even for defects outside the probed area. This effect is reduced through increasing the contact resistance between the electrode and the wafer or the sheet resistance of the electrode. Although this work focused on silicon heterojunction technology, the studied phenomena also apply to devices using high-temperature passivated contacts, for which lower contact resistance values are reported.

The authors acknowledge Nicolas Badel, Patrick Wyss and Christophe Allebé for the high-quality wet-processing and metallisation. We also thank Vincent Paratte, Cédric Bucher and Aymeric Schafflützel for technical support. This project is funded by the Qatar Foundation and the Swiss National Science Foundation under the Ambizione Energy Grant ICONS (PZENP2_173627).